\documentclass[preprint,showpacs,preprintnumbers,amsmath,amssymb]{revtex4}
\usepackage{graphicx}
\usepackage{slashed}
\usepackage{epstopdf}

\begin{document}

\title{Scalar Dark Matter in light of LEP and ILC Experiments}

\author{F.~Rossi-Torres}%
\email{ftorres@ifi.unicamp.br}
\affiliation{Instituto de F\'isica Gleb Wataghin, Universidade Estadual de Campinas - UNICAMP, Rua S\'ergio Buarque de Holanda, 777, 13083-859, Campinas-SP, Brazil}
\author{C.~A.~Moura}%
\email{celio.moura@ufabc.edu.br}
\affiliation{Centro de Ci\^encias Naturais e Humanas, Universidade Federal do ABC - UFABC, Av. dos Estados, 5001, 09210-580, Santo Andr\'e-SP, Brazil}

\date{\today}

\begin{abstract}
In this work we investigate a scalar field dark matter model with mass in the order of 100 MeV.
We assume dark matter is produced in the process $e^-+e^+\to \phi +\phi^*+\gamma$, 
that, in fact, could be a background for the standard process $e^-+e^+\to \nu +\bar\nu+\gamma$ extensively studied at LEP. We constrain the chiral couplings, $C_L$ and $C_R$, 
of the dark matter with electrons through an intermediate fermion of mass $m_F=100$~GeV and 
obtain $C_L=0.1(0.25)$ and $C_R=0.25(0.1)$ for the best fit point of our $\chi^2$ analysis.
We also analyze the potential of ILC to detect this scalar dark matter for two configurations: ({\it i}) center of mass energy $\sqrt{s}=500$~GeV and luminosity $\mathcal{L}=250$~fb$^{-1}$, and ({\it ii}) center of mass energy $\sqrt{s}=1$~TeV and luminosity $\mathcal{L}=500$~fb$^{-1}$. The differences of polarized beams are also explored to better understand the chiral couplings.
\end{abstract}


\maketitle

\section{Introduction}

Presently, many different sources of data point to the existence of dark matter. In the standard cosmological model, dark matter contributes to the total universe energy budget with an energy density of approximately 23.5\%~\cite{komatsu}. 
Although we do not know its nature, according to our present knowledge, dark matter must be a new kind of neutral and stable particle~\cite{review1,review2,review3}. Probably, the most well-known kind of hypothesized dark matter particle is the Weakly Interacting Massive Particle (WIMP), which is constructed mainly by supersymmetric models. 
Although several attempts and experimental proposals have been made to detect WIMPs, their existence is not yet confirmed.
The observation of a dark matter candidate could be realized once it scatters in nuclei that compound a solid state detector~\cite{direct}. Examples of this kind of direct detection experiment are: XENON~\cite{xenonexp}, DAMA~\cite{dama}, CoGeNT~\cite{cogent} and CDMS~\cite{cdms}.

However, dark matter could be detected in indirect ways as, for example, annihilation in ({\it i}) gamma rays, such as the ones detected in FermiLAT experiment~\cite{fermilat},
({\it ii}) charged particles, explored by
PAMELA~\cite{pamela}, ({\it iii}) neutrinos, searched for by IceCube~\cite{icecube}, etc. For a review of indirect dark matter search, see~\cite{review4}. 
Another way to search for dark matter particles is to possibly produce them using colliders. LHC is an example of a hadronic collider while the past LEP~\cite{aleph1,aleph2,aleph3,delphi,l31,l32,l33,opal1,opal2,opal3} at CERN or the future International Linear Collider (ILC)~\cite{ilc} are examples of leptonic colliders.

Dark matter models have been explored using LHC data, such as in~\cite{lhc1,lhc2,lhc3,lhc4}.
This search is mainly based on the dark matter missing energy plus the observed final states from the Higgs decay products. In LEP/ILC, the physical strategy is similar: the annihilation of particles, in this case an electron and a positron, into a pair of dark matter candidates, which are invisible~\cite{colliders1}. However, this production can be followed also by a photon that can be detected.
In LEP, an excess of events related with dark matter plus mono-photon production has not been found beyond the expected background, and limits to such an interaction were placed instead~\cite{lep1,lep2}. In contrast to previous analyses, we include in our work a dark matter model which takes into account different couplings with right- and left-handed fermions.

Despite the aforementioned experimental efforts, WIMP dark matter remains a hypothesis and its existence is still an ongoing search at the LHC~\cite{bertonenature}. 
It is expected that the mass scale of a WIMP dark matter is of the order of 100~GeV.
However, authors in Ref.~\cite{xenon} claim that it is possible to have dark matter candidates with masses well below the GeV scale.
Other works considered the dark matter particle to be scalar singlet fields~\cite{scalar1,scalar2,scalar3,scalar4,scalar5} or more complex models, with different symmetries and other constructions, as explored in~\cite{otherdmmodel1,otherdmmodel2,otherdmmodel3,otherdmmodel4,otherdmmodel5}.

In this study, we analyze a scalar dark matter particle explored by Boehm and Fayet~\cite{boehm} that has a mass lighter than $\mathcal O$(1~GeV). It might be produced in the annihilation of electrons and positrons in the leptonic colliders cited above: LEP and the future ILC. The scalar dark matter model under consideration has four main parameters: {\it i}) one for the coupling with left chirality leptons, ({\it ii}) one for the coupling with right chirality leptons, ({\it iii}) the dark matter mass, and ({\it iv}) the mass of a heavy intermediate fermion. We find the constraints in the coupling constants for different intermediate fermion masses.
There is no sensitivity to the dark matter mass because the collision energy is much higher than the dark matter mass itself.

This article is organized as follows: In section~\ref{crsection} we describe the cross section and summarize the model that we test. In section~\ref{results} we present our results for the LEP data (\ref{lep}) and make predictions for the ILC (\ref{ilc}). This section also presents discussion about the results we obtain. In section~\ref{conclusions} we conclude our work. 

\section{Photons plus invisible energy in $e^+e^-$ coliders}\label{crsection}
\begin{figure}[h!] 
\includegraphics[scale=1.0]{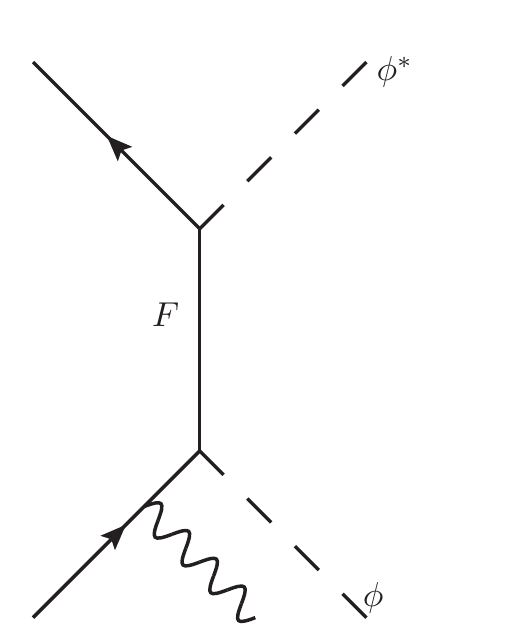}
\includegraphics[scale=1.0]{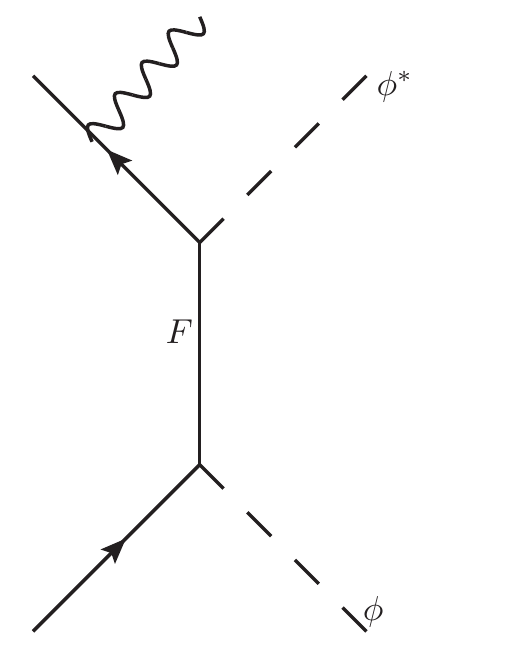}
\includegraphics[scale=1.0]{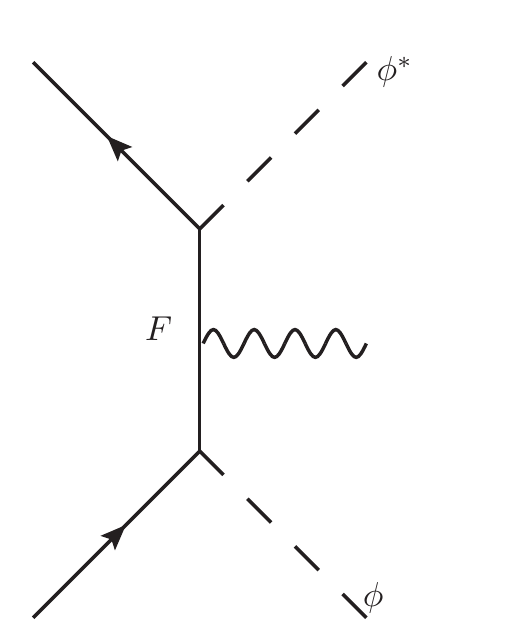}
\caption{Relevant Feynman diagrams for the process $e^- + e^+ \to \phi + \phi^* + \gamma$.}
\label{feynman}
\end{figure}

Consider the interaction
\begin{equation}
e^- + e^+ \to \phi+ \phi^* + \gamma, 
\end{equation}
represented by the Feynman diagrams in Fig.~\ref{feynman}, where $\phi$ is the scalar dark matter and $\phi^*$ is its conjugate with $\phi \ne \phi^*$.
These scalar dark matter particles couple to standard model
fermions and to a nonstandard intermediate fermion ($F$). 
The mass of the fermion $F$ mediating the interaction is typically above $\approx 100$~GeV. This is a reasonable assumption, since it is compatible with the non detection up to now of new charged and heavy fermions. This new nonstandard fermion field may be some mirror partner of other fermions that we know in our universe. The relevant Feynman rules in our case are expressed by $\phi(C_L\bar f_L F_R + C_R \bar f_R F_L +h.c.)$, where $\phi$ is the scalar field dark matter, and $C_L$ and $C_R$ are the Yukawa couplings, respectively to the left-handed and right-handed standard model fermions. In our analysis, these couplings are free parameters, together with the dark matter mass, $m_\phi$, and the mass of the intermediate fermion, $m_F$. See~\cite{boehm} for more details of the model building. 

The cross section of the $e^-+e^+ \to \phi + \phi^* + \gamma$ process can be evaluated at tree level using the ``radiator approximation''~\cite{radiator}. We present the associated cross section as
\begin{equation}
\sigma(s)=\int dx \int dc_\gamma H(x,s_\gamma;s)\sigma_0(\hat s),
\label{rad}
\end{equation}
where $s$ is the square of the center of mass energy, $x=2E_\gamma/\sqrt{s}$, $E_\gamma$ is the emitted photon energy. The cross section $\sigma_0$ is the cross section associated with the dark matter production by electron-positron annihilation, $e^-+e^+ \to \phi + \phi^*$, written in terms of the parametrized $\hat s=s(1-x)$. The total cross section, $\sigma_0$, is related to the following differential cross section: 
\begin{equation}
\frac{d\sigma_0}{d\Omega}=\left(\frac{1}{8\pi}\right)^2 \frac{|M|^2}{2s}\frac{\sqrt{s/4-m_\phi^2}}{\sqrt{s/4}}\,,
\label{diff_cross}
\end{equation}
where $\Omega$ is the solid angle and $|M|^2$ is the square amplitude evaluated considering the Feynman diagrams in Fig.~\ref{feynman}.
The radiator function $H$ is described in the following equation:
\begin{equation}
H(x,s_\gamma;s)=\frac{2\alpha}{\pi x s_\gamma}\left[\left(1-\frac{x}{2} \right)^2+\frac{x^2 c_\gamma^2}{4} \right]\,,
\end{equation}
for $c_\gamma=\cos\theta_\gamma$ and $s_\gamma=\sin\theta_\gamma$\,, where $\theta_\gamma$ is the photon emission angle.

The radiator function is a good approximation when the emitted photon is neither soft, i.e., with high transverse momentum, nor collinear to the incoming $e^-$ or $e^+$. It is important to emphasize that this approximation does not depend on the nature of the electrically neutral particles produced with the photon. This is a reasonable approximation and it works very well for our evaluations. In figure~2 of Ref.~\cite{backilc1}, we find the comparison of the analytical solution and the radiator approximation. It is a very good approximation up to $E_\gamma\approx450$~GeV ($\sqrt{s}\approx1$~TeV), which is our most powerful configuration for ILC (Sec.\ref{ilc}).

\begin{figure}[!h]
\centering
\includegraphics[scale=0.45]{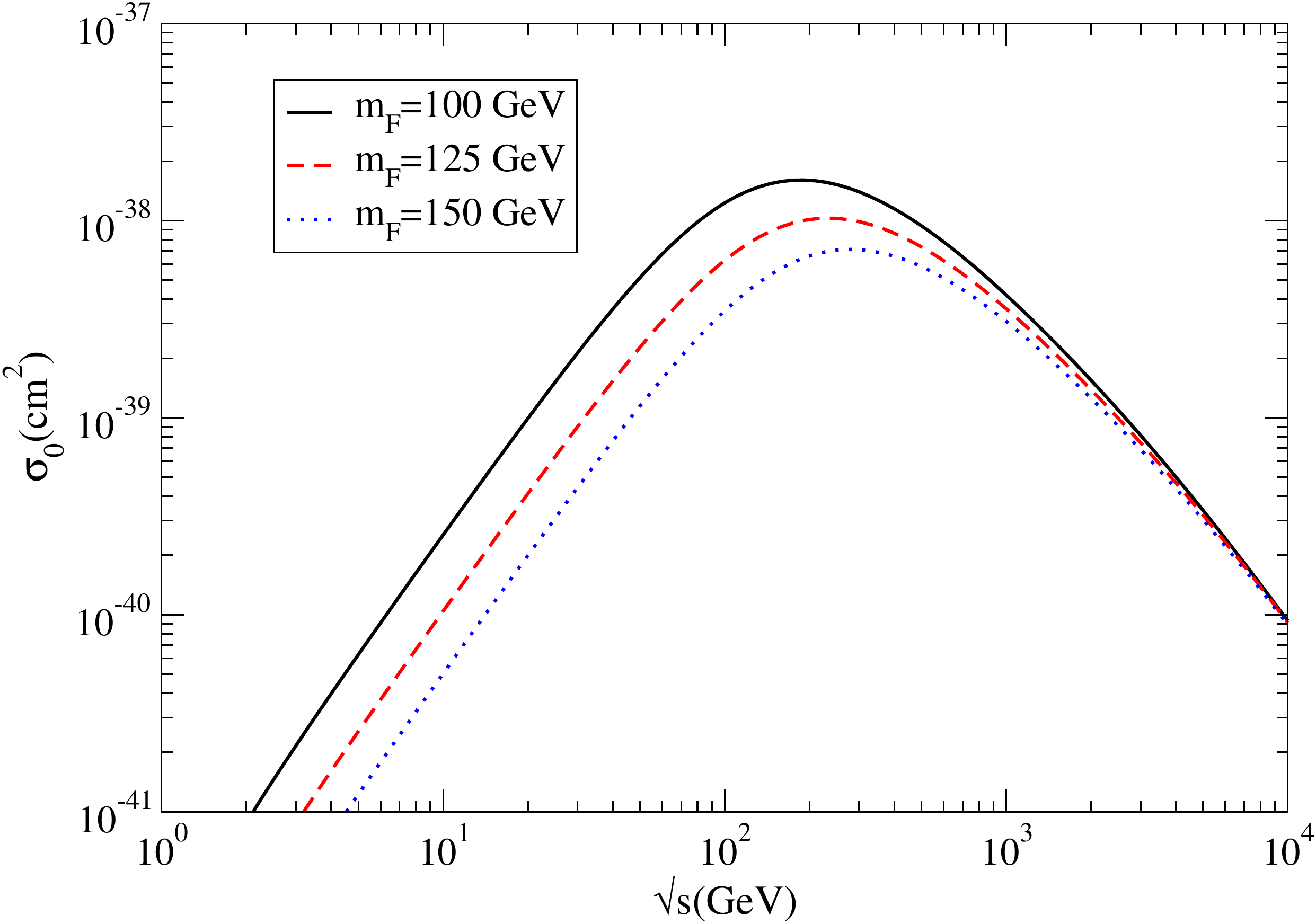}
\caption{({\it Color online}) Cross section for the process $e^- + e^+ \to \phi + \phi^*$, considering three different fermion masses. The scalar dark matter mass $m_\phi=100$~MeV and $C_L=C_R=0.1$.}
\label{fig1}
\end{figure}

We calculate the amplitude for the process $e^-+e^+ \to \phi + \phi^*$, since it can provide information on how relevant the missing energy process can be when compared to the neutrino's missing energy on $e^-+e^+\to \nu+\bar\nu$. Our evaluation takes into account that $m_F>>m_\phi>>m_e\,.$ In Fig.~\ref{fig1}, for different values of $m_F$, $C_L=C_R=0.1$, and $m_\phi=100$~MeV, we have the total cross section for this process in terms of the center of mass energy of the collision, $\sqrt{s}$\,.
From XENON10 data, authors of Ref.~\cite{xenon} obtained the strongest bound on the scattering cross section between dark matter and electrons at a 100~MeV dark matter mass. Although XENON is a direct detection experiment, this bound at 100~MeV dark matter mass was a motivation for us to use this value for our scalar dark matter mass.

In order to compare the cross sections ($\sigma_0$) of $e^-+e^+\to \phi+\phi^*$ with the Standard Model process $e^-+e^+\to \nu+\bar\nu$ ($\sigma_{SM}$), we consider $\sqrt{s}=100$~GeV as an example. In~Fig.~\ref{fig1} we illustrate that, for $m_F=100$~GeV, $C_L=C_R=0.1$, and $m_\phi=100$~MeV, then $\sigma_0 \approx  10^{-38}$~cm$^2$. Using the cross section calculated in~\cite{thooft,dicus}, $\sigma_{SM}\approx 10^{-35}$~cm$^2$. So we are clearly describing a subleading process if compared with the electron-positron annihilation into neutrinos.

\section{The process $e^+e^-\to\gamma\phi\phi^*$ at LEP and ILC}\label{results}

We divide this section in two: first, in~\ref{lep}, we present our results using data from LEP; then, in~\ref{ilc}, we present 
the ILC potential to investigate dark matter in light of monophoton production from $e^+e^-$ collisions.

\subsection{LEP Results}\label{lep}

We analyse the possible existence of low mass dark matter, $\mathcal O$(100)~MeV, using LEP data from the experiments~\cite{aleph1,aleph2,aleph3,delphi,l31,l32,l33,opal1,opal2,opal3}.
The center of mass energy $\sqrt{s}$ of the $e^+e^-$ collision varies from 130~GeV to 207~GeV and the luminosity varies from 2.3~pb$^{-1}$ to 173.6~pb$^{-1}$.

\begin{figure}[!h]
\centering
\includegraphics[scale=0.9]{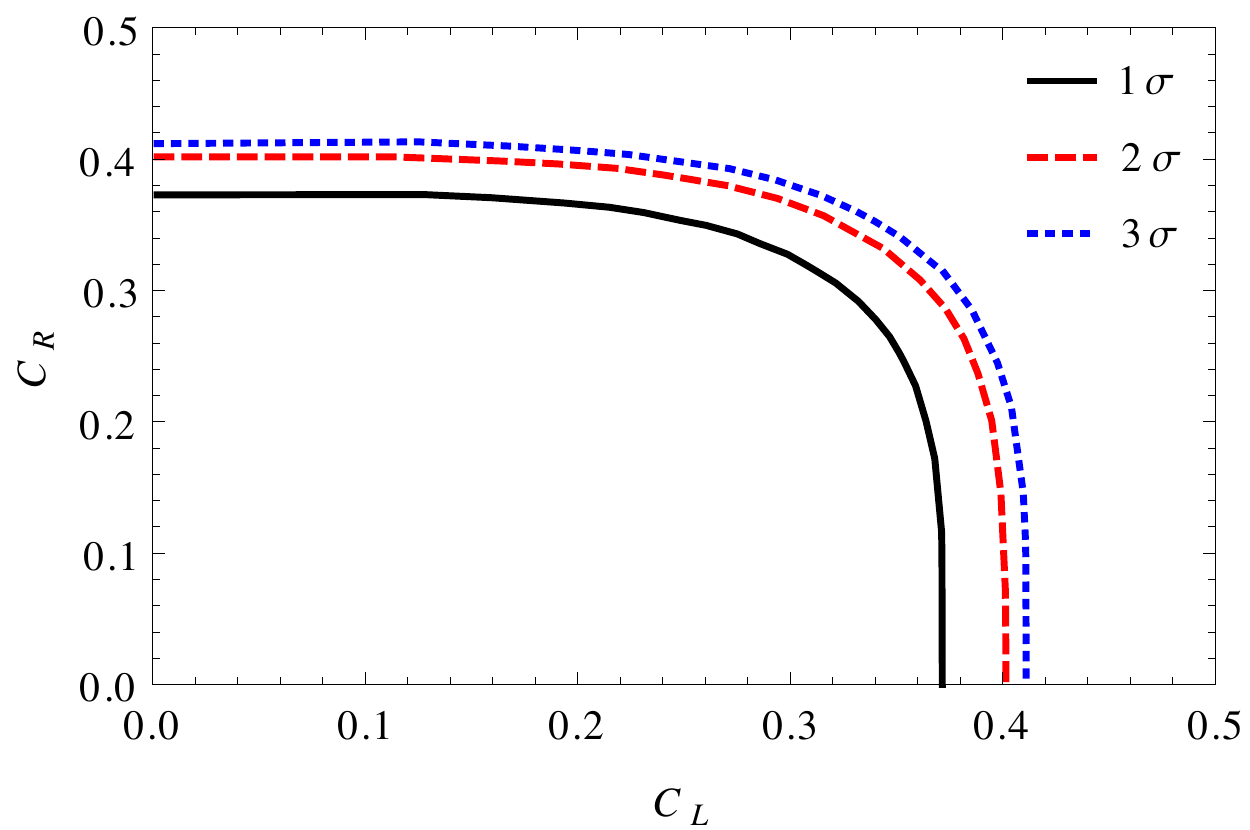}
\caption{({\it Color online}) Allowed couplings in the parameter space $C_L-C_R$ for 68\% (black curves), 90\% (red dashed curves), and 95\% (blue dotted curves) CL. $m_F=100$~GeV and a $m_\phi=100$~MeV. The best fit point occurs for $C_L=0.1$ and $C_R=0.25$ or $C_L=0.25$ and $C_R=0.1$.}
\label{fig2}
\end{figure}

We present in Fig.~\ref{fig2} the allowed regions in the $C_L-C_R$ space of parameters for
the confidence level CL = 68\% (black curves), 90\% (red dashed curves), and 95\% (blue dotted curves), 
for $m_F=100$~GeV, and the dark matter mass $m_\phi=100$~MeV. 
We find that the $\chi^2$ value increases with $m_F$. On the other hand, there is no significant variation of the $\chi^2$ value when $m_\phi$ changes. Actually, the lack of sensitivity 
with respect to $m_\phi$
is related to the fact that such a mass is orders of magnitude below the 
experiment energy scale, $\sqrt{s}$, which is of order of 100~GeV. The best-fit point 
for $m_F=100$~GeV is at
$C_L=0.1$ and $C_R=0.25$ or at $C_L=0.25$ and $C_R=0.1$, with $\chi^2_{min}=21.97$.

The fact that there is no sensitivity on constraining the dark matter mass with values less than 1~GeV, considering LEP data, was also noticed in~\cite{lep2}. It is worth noting that the bounds we obtain are compatible with the limits of annihilation cross section of dark matter candidates that generate the known dark matter abundances. According to~\cite{boehm}, if one considers that $\Omega_{\phi} h^2\approx 0.1$, this can be achieved by $C_L C_R \approx 0.01-0.1$ and $m_F\approx 100 - 1000~\mbox{GeV}$, for any value of $m_{\phi}$ less than $\mathcal O$(1~GeV).

\subsection{ILC Predictions}\label{ilc}

Identifying the process $e^-+e^+\to \phi +\phi^* +\gamma$ is difficult because of the background from the radiative neutrino production $e^-+e^+\to \nu +\bar\nu+ \gamma$. The International Linear Collider (ILC)~\cite{ilc} is expected to have a much higher luminosity and beam polarization than LEP. This provides a higher sensitivity to distinguish a process of interest from all possible backgrounds. For instance, it has been shown for a WIMP dark matter, with an integrated luminosity of 500~fb$^{-1}$, that cross sections as small as 12~fb can be observed at the 5$\sigma$ level, considering only the statistical uncertainty and fully polarized beams~\cite{ilcwimp}. Barthels {\it et al.}~\cite{barthels} show that, with the luminosity of 500~fb$^{-1}$, it is possible to infer the helicity structure of the interaction involved, and the masses and cross sections can be measured with a relative accuracy of the order of 1\%.

In order to avoid the collinear and infrared divergences, we impose that $E_\gamma>8$~GeV and $-0.995<\cos\theta_\gamma<0.995$; also, we consider $E_\gamma\le 220$~GeV when $\sqrt{s}=500$~GeV and $E_\gamma\le 450$~GeV for $\sqrt{s}=1$~TeV, as it is assumed, e.g., in~\cite{dipole_ilc}. These cuts are safe cuts, avoiding a higher background contamination, since there are $Z$-resonances $\sqrt{s}/2(1-M_z^2/s)$ for the process $e^-+e^+\to \nu+\bar\nu+\gamma$.

As already emphasized, $e^-+e^+\to \nu+\bar\nu+\gamma$ (bg1) is the main contamination channel and its number of events depends on the beam polarization. The second relevant channel of the so-called neutrino background is $e^-+e^+\to \nu+\bar\nu+\gamma+\gamma$ (bg2), where there can be an emission of a second photon, which is not detected. This background channel contributes to about 10\% of the total neutrino background. Finally, there is also the Bhabha scattering of leptons with an emission of a photon: $e^-+e^+\to e^-+e^++\gamma$ (bg3). This process can contribute in almost the same amount as the neutrino background and it is mostly polarization independent. In Table~\ref{tabback} we show the number of all relevant background events, taking into account three possible configurations of the beam polarization: $i)$ Unpolarized, i.e., $(P_{e^-},P_{e^+})=(0.0,0.0)$, $ii)$ $(P_{e^-},P_{e^+})=(+0.8,-0.3)$, and $iii)$ $(P_{e^-},P_{e^+})=(-0.8,+0.3)$. All the backgrounds are estimated in~\cite{backilc1} for a luminosity of 1~fb$^{-1}$ and $\sqrt{s}=500$~GeV. Numbers in parentheses in Table~\ref{tabback} are the number of background events for the same luminosity, but for $\sqrt{s}=1$~TeV. 

\begin{table}[h!]
\centering
    \begin{tabular}{| l | l | l | l |}
    \hline
     $(P_{e^-},P_{e^+})$ & $bg1$ &  $bg2$ &  $bg3$ \\ 
    \hline
    \hline
    	(0,0) & 2257 (2677) & 226 (268) & 1218 (304) \\
    \hline
        (+0.8,-0.3) & 493 (421) & 49 (42) & 1218 (304) \\
    \hline 
	(-0.8,+0.3) & 5104 (6217) & 510 (622) & 1218 (304) \\
   \hline 
    \end{tabular}
\caption{Number of background events from the three different main
channels: $e^-e^+\to \nu+\bar\nu+\gamma$ (bg1); $e^-+e^+\to \nu+\bar\nu+\gamma+\gamma$ (bg2); and $e^-+e^+\to e^-+e^++\gamma$ (bg3). The numbers are given for an integrated luminosity of 1~fb$^{-1}$ and $\sqrt{s}=500$~GeV(1~TeV), considering three different beam polarizations: $(P_{e^-},P_{e^+})=(0,0)$ (unpolarized), $(P_{e^-},P_{e^+})=(+0.8,-0.3)$ and $(P_{e^-},P_{e^+})=(-0.8,+0.3)$. From~\cite{backilc1}.}
\label{tabback}
\end{table}

The number of background events is modified by the polarization of the beam. This is an important consideration that improves the study of dark matter models at ILC. The cross section for the dark matter production, $\sigma_0$, evaluated using Eq.~(\ref{diff_cross}),
is also affected. When we consider polarization the cross section ($\sigma_{pol}$) is given by~\cite{colliders1}
\begin{eqnarray} \nonumber
\sigma_{pol} & = & \frac{1}{4}(1+P_-)\left[(1+P_+)\sigma_0(e_R^- e_L^+)+(1-P_+)\sigma_0(e_R^- e_R^+) \right] + \\
         &  +  & \frac{1}{4}(1-P_-)\left[(1+P_+)\sigma_0(e_L^- e_L^+)+(1-P_+)\sigma_0(e_L^- e_R^+) \right]\,,
\label{pol}
\end{eqnarray}
where $P_-$ and $P_+$ are the electron and positron polarizations, respectively. $P_i=0$ ($i=\pm$) represents unpolarized beams and $P_i=1$ represents pure right-handed ($i=-$) electron beam and left-handed ($i=+$) positron beam. The $\sigma_0(e_j^- e_k^+)$, $j,k=L,R$, are the cross sections for the different states, left or right, of the electron and positron polarizations in the beam and are evaluated using Eq.~(\ref{diff_cross}) and the amplitudes of the Feynman diagrams in Fig.~\ref{feynman}. For our process of interest, $e^- + e^+ \to \phi+\phi^*+\gamma$, in the polarized case, we take the value of $\sigma_{pol}$ now calculated by Eq.~(\ref{pol}) and insert in Eq.~(\ref{rad}) substituting $\sigma_0$. The procedure of cross section evaluation is very similar to the unpolarized case.  

The constraints on the parameters of the model can be evaluated by~\cite{upper1,upper2,upper3}
\begin{equation}
N_{sig}+N_{bg}-A\sqrt{N_{sig}+N_{bg}}>N_{bg}+A\sqrt{N_{bg}},
\end{equation}
where $A=1.64$ for 95\% of confidence level, $N_{sig}$ and $N_{bg}$ are, respectively, the number of signal events and background events after considering all the cuts.

\begin{figure}[!h]
\centering
\includegraphics[scale=0.9]{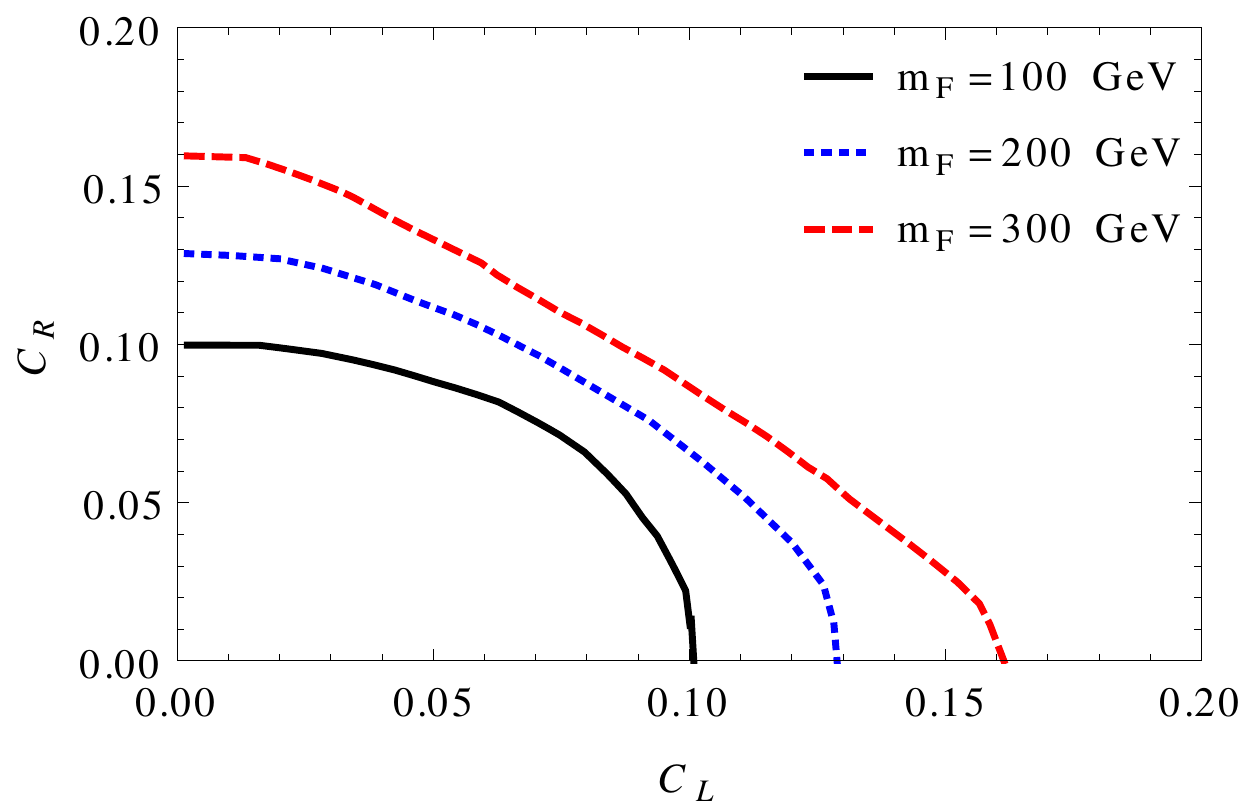}
\caption{({\it Color online}) Bounds at 95\% CL on the $C_L$-$C_R$ couplings considering $m_\phi=100$~MeV, $\sqrt{s}=500$~GeV and $\mathcal{L}=250$~fb$^{-1}$. Black solid curve is for $m_F=100$~GeV; blue dotted line is for $m_F=200$~GeV; and red dashed line is for $m_F=300$~GeV.}	
\label{fig3}
\end{figure}

Fig.~\ref{fig3} illustrates the bounds for $C_L$ and $C_R$ at 95\% CL, considering $m_\phi=100$~MeV, $\sqrt{s}=500$~GeV and $\mathcal{L}=250$~fb$^{-1}$, for three different values of $m_F$, i.e., black solid line is for $m_F=100$~GeV; blue dotted line is for $m_F=200$~GeV; and red dashed line is for $m_F=300$~GeV. These values are evaluated for an unpolarized beam.

\begin{figure}[!h]
\centering
\includegraphics[scale=0.8]{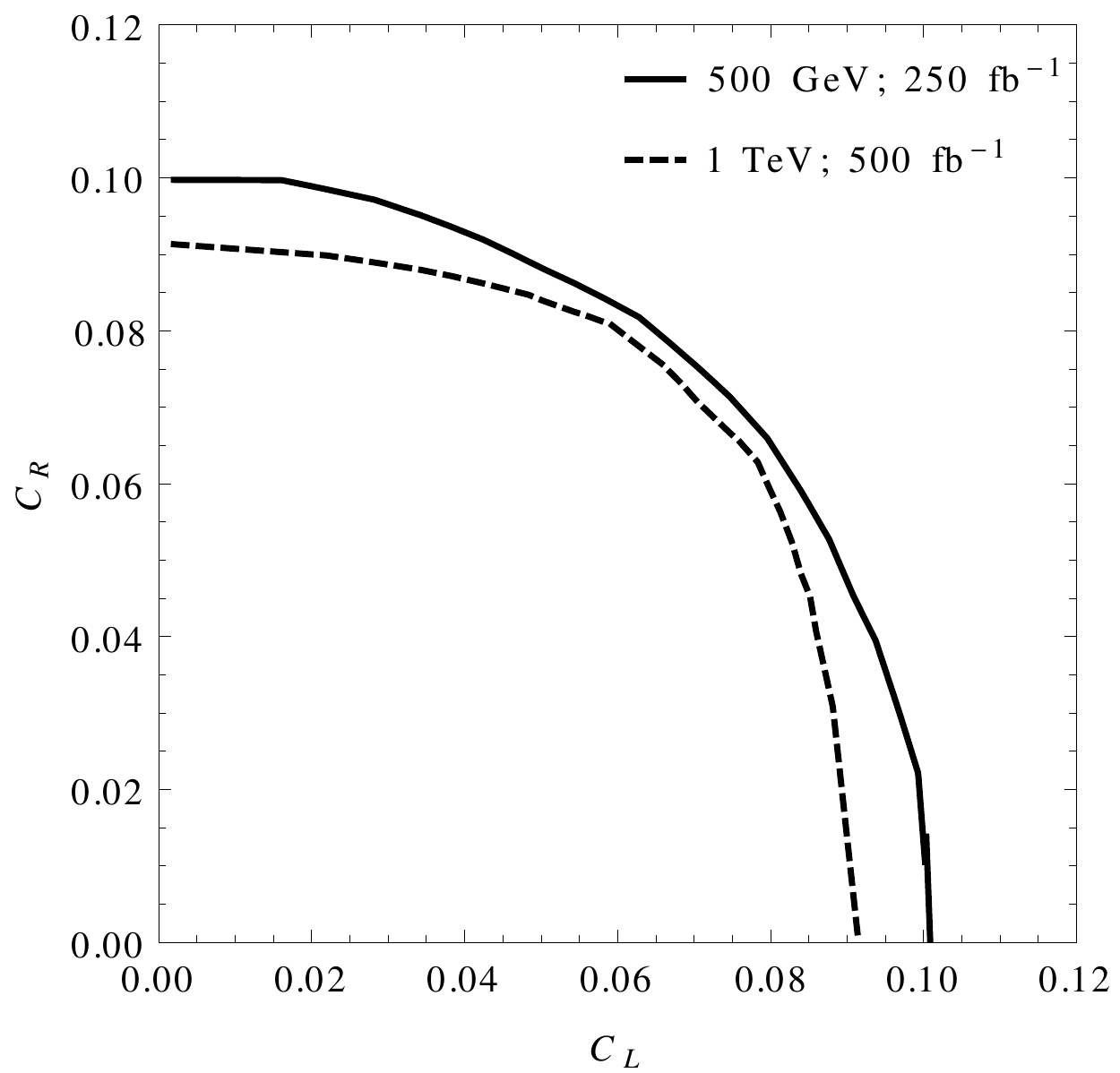}
\caption{Bounds at 95\% CL on the $C_L$-$C_R$ couplings considering $m_\phi=100$~MeV and $m_F=100$~GeV. Solid curve is for ILC with $\sqrt{s}=500$~GeV and $\mathcal{L}=250$~fb$^{-1}$; Dashed curve is for ILC with $\sqrt{s}=1$~TeV and $\mathcal{L}=500$~fb$^{-1}$.}
\label{fig4}
\end{figure}

As in the LEP case, a variation in dark matter mass does not affect the constraints, since $\sqrt{s}>>m_\phi$. We found that, for $m_F=100$~GeV, we obtain better constraints, as expected with the increased luminosity. ILC can provide, roughly speaking, coupling constraints four times stricter than LEP in similar conditions.
For an ILC configuration of $\sqrt{s}=1$~TeV and $\mathcal{L}=500$~fb$^{-1}$ (dashed curve in Fig.~\ref{fig4}), $C_L$-$C_R$ coupling parameters are more constrained than for the configuration of $\sqrt{s}=500$~GeV and $\mathcal{L}=250$~fb$^{-1}$ (solid curve). Both curves were obtained considering $m_\phi=100$~MeV and $m_F=100$~GeV. We observe this behavior, when the energy of the collision as well as the luminosity of the experiment increases, due to a slight reduction of the background events and an increase of events in the dark matter channel production (Table~\ref{tabback}). Although the cross section for the process is reduced when the collision energy increases, this reduction is not significant when compared with the increase in the luminosity.

\begin{figure}[!h]
\centering
\includegraphics[scale=0.8]{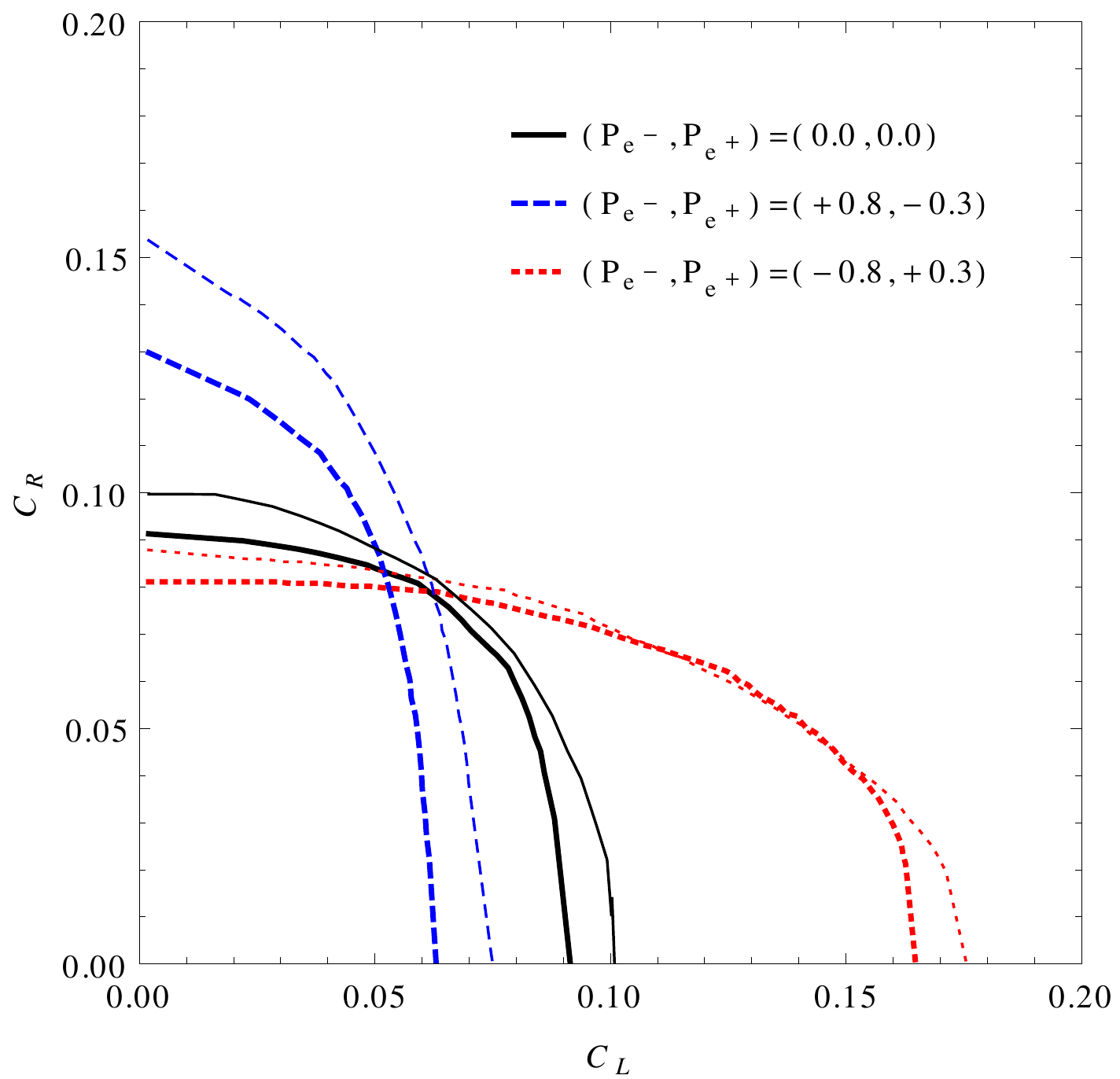}
\caption{({\it Color online})Bounds at 95\% CL on the $C_L$-$C_R$ couplings considering $m_\phi=100$~MeV and $m_F=100$~GeV. ILC configuration has been set for $\sqrt{s}=500$~GeV and $\mathcal{L}=250$~fb$^{-1}$. Black solid line is for unpolarized beam; blue dashed line is for $(P_{e^-},P_{e^+})=(+0.8,-0.3)$; and red dotted line is for $(P_{e^-},P_{e^+})=(-0.8,+0.3)$.  Thinner lines for $\sqrt{s}=500$~GeV and $\mathcal{L}=250$~fb$^{-1}$ and thicker lines for $\sqrt{s}=1$~TeV and $\mathcal{L}=500$~fb$^{-1}$.}
\label{fig5}
\end{figure}

We also calculated the results by taking into account two configurations for the polarization of the beam: $(P_{e^-},P_{e^+})=(+0.8,-0.3)$ and $(P_{e^-},P_{e^+})=(-0.8,+0.3)$. Fig.~\ref{fig5} represents the 95\% CL on the $C_L$-$C_R$ couplings considering $m_\phi=100$~MeV, $m_F=100$~GeV. The black solid line is for unpolarized beam; the blue dashed line is for $(P_{e^-},P_{e^+})=(+0.8,-0.3)$; and the red dotted line is for $(P_{e^-},P_{e^+})=(-0.8,+0.3)$. The thinner (upper) lines represent the ILC configuration of 
$\sqrt{s}=500$~GeV and $\mathcal{L}=250$~fb$^{-1}$ and the thicker lines exemplify the ILC configuration of $\sqrt{s}=1$~TeV and $\mathcal{L}=500$~fb$^{-1}$.

As already mentioned and presented in Fig.~\ref{fig4}, greater luminosity in the experiment imposes a more constrained region of the space of parameters. As expected, when we consider the beam polarization, asymmetries in the curves appear. As noticed in Fig.~\ref{fig5}, the polarization configuration of the blue dashed line is almost the rotate red dotted curve, since there is an inversion of the beam polarization: $(P_{e^-},P_{e^+})=(+0.8,-0.3) \to (P_{e^-},P_{e^+})=(-0.8,+0.3)$. Differences between these two curves are related to the number of background events when we modify the polarization, as represented in Tab.~\ref{tabback}. 

\section{Conclusions}\label{conclusions}

In this work, we analyzed data from LEP in the context of a scalar dark matter model. The production of the dark matter particle, $e^-+e^+\to \phi +\phi^*+\gamma$, is confronted with other standard model backgrounds, such as $e^-+e^+\to \nu +\bar\nu+\gamma$. We constrain the $C_L$ and $C_R$ couplings for an intermediate heavy fermion mass $m_F=100$~GeV and dark matter mass $m_\phi=100$~MeV. We obtain $C_L=0.1(0.25)$ and $C_R=0.25(0.1)$ as best-fit points. When $m_F$ increases, $\chi^2$ becomes inaccurate and unresponsive to the dark matter mass ($m_\phi$). We also investigate the potential of ILC to constrain scalar field dark matter models. Using an unpolarized beam and $E_\gamma\le 220$~GeV when we consider $\sqrt{s}=500$~GeV and $\mathcal{L}=250$~fb$^{-1}$, and $E_\gamma\le 450$~GeV for $\sqrt{s}=1$~TeV and $\mathcal{L}=500$~fb$^{-1}$, it is clear that ILC is more sensitive than LEP to investigate scalar field dark matter models, since it will have a greater luminosity, polarization information, and an improved comprehension of the backgrounds.

Although ILC has no sensitivity for $m_\phi$, since $\sqrt{s}>>m_\phi$, it can have different polarization configurations. There is great potential to explore models where there are distinctions in couplings between dark matter with left- and righ-handed fermions, as seen in Fig.~\ref{fig5}.   
Our study signalizes the importance to deeply explore dark matter and broken chiral symmetry models and the potential to accomplish this at the future ILC or any other future electron-positron collider. It opens the possibility to study models that contain a large spectrum of dark matter masses and to explore nonstandard weak couplings, such as found in~\cite{ufabc_exotico,exotico2,exotico3,exotico4}, among many other works on the subject. Lepton colliders are very important tests for dark matter, since backgrounds can, in principle, be much better understood and provide clearer event signals.

\vspace{1cm}

{\bf Acknowledgements}: F.~Rossi-Torres would like to thank Conselho Nacional de Desenvolvimento Cient\'ifico e Tecnol\'ogico (CNPq) for financial support (grant number 150102/2013-5). The work of C.A.M. was partially supported by Funda\c{c}\~ao de Amparo \`a Pesquisa do Estado de S\~ao Paulo (FAPESP), under the grant 2013/22079-8. The authors would like to thank O.~L.~G.~Peres for the suggestion of the subject we have developed here.

\end{document}